\documentclass[journal,compsoc]{IEEEtran}
\usepackage[T1]{fontenc}
\usepackage[utf8]{inputenc}
\usepackage{newtxtext,newtxmath}
\usepackage{graphicx}
\usepackage{amsmath}
\usepackage{etoolbox}
\usepackage{cite}
\usepackage{booktabs}
\usepackage{xurl}
\usepackage{hyperref}
\hypersetup{hidelinks}
\title{PSD-GRN: Pseudotime-Enhanced Signed Directed Graph Learning for Gene Regulatory Network Inference}
\author{Qi~Chen, Kai~Wang, Hongliang~Zhang, Wei~Xiong, Yuan-nong~Ye, and Bin~Zhao%
\IEEEcompsocitemizethanks{%
\IEEEcompsocthanksitem This work was supported by the Guizhou Provincial Basic Research Program (Natural Science Category) Project ZK [2023]298, Grant Qiankehe Foundation MS[2026]543 (General Program), and the National Natural Science Foundation of China (No. 12461067).
\IEEEcompsocthanksitem Q. Chen, K. Wang, Y. Ye, and B. Zhao are with the School of Biology and Engineering, Guizhou Medical University, Guiyang 561113, China.
\IEEEcompsocthanksitem Q. Chen, H. Zhang, and W. Xiong are with the School of Mathematics and System Science, Xinjiang University, Urumqi 830017, China.
\IEEEcompsocthanksitem Q. Chen and K. Wang contributed equally. Corresponding authors: B. Zhao (scott84@sina.com) and Y. Ye (ynye.uestc@gmail.com).}%
}

\makeatletter
\expandafter\patchcmd\csname\string\IEEEbiography\endcsname{\vskip \@IEEEBIOskipN plus 1fil minus 0\baselineskip}{\vskip 1.5\baselineskip}{\typeout{BIOGRAPHY_SPACING_PATCH_PASS}}{\PackageError{PSD-GRN}{Biography spacing patch failed}{Check IEEEtran version}}
\makeatother

\begin{document}
\maketitle
\begin{abstract}
A complete characterization of a gene regulatory relationship requires determining its existence, direction, and activating or repressive effect. Existing single-cell gene regulatory network (GRN) inference methods often rely on static expression features, model direction and sign separately, and restrict candidate regulators to predefined transcription factors (TFs). Incomplete reference networks and TF catalogues may therefore exclude potential regulators. We propose PSD-GRN, a pseudotime-enhanced signed directed graph learning framework that partitions cells into consecutive pseudotime stages, extracts stage-specific expression features, and sequentially integrates gene representations. A prior graph is constructed from partially observed regulatory relationships, and magnetic Laplacian spectral convolution encodes regulatory direction and activating or repressive effects. PSD-GRN evaluates ordered gene pairs within the selected gene set and identifies potential regulatory sources beyond predefined TF lists. It supports both directed edge-existence prediction and signed directed regulatory relationship prediction. Across single-cell RNA-seq benchmarks, PSD-GRN achieves comparable or superior performance to baseline methods. In an idiopathic pulmonary fibrosis (IPF) case study, PSD-GRN identifies candidate gains, losses, and sign changes in regulatory relationships across repair- and disease-associated epithelial trajectories, with some changes reproduced in an independent cohort. These findings support its use in identifying potential regulators and trajectory-associated regulatory changes.
\end{abstract}
\begin{IEEEkeywords}
Single-cell RNA sequencing, Gene regulatory network inference, Pseudotime, Signed directed graph, Magnetic Laplacian, Pulmonary fibrosis
\end{IEEEkeywords}
\section{Introduction}\label{introduction}

Gene regulatory networks (GRNs) describe interactions through which regulators influence target genes, shaping cell differentiation, tissue development, and disease-associated cellular changes \cite{ref1}. Single-cell RNA sequencing (scRNA-seq) captures transcriptional heterogeneity across cells, while trajectory inference organizes cellular states along developmental or disease-associated transitions. For GRN inference, these data provide information about expression variation across cellular states, but the direction and activating or repressing effects of candidate regulatory relationships remain difficult to determine. Reference GRNs provide complementary regulatory information, although their annotations are incomplete. Integrating pseudotime-resolved expression variation with signed directed regulatory topology can therefore support the prediction of regulatory states for ordered gene pairs. These predictions can also be used to compare candidate regulatory changes across trajectories associated with distinct cellular outcomes.

Methods for single-cell GRN inference use different sources of information, including gene-expression dependencies, inferred pseudotime, and regulatory network topology. Regression and information-theoretic approaches estimate candidate interactions from gene expression, while pseudotime-based methods use inferred cell orderings to characterize expression changes and estimate regulatory relationships \cite{ref2,ref3,ref4}. Deep learning methods capture nonlinear gene-pair relationships or incorporate prior regulatory graphs to predict unrecorded interactions \cite{ref5,ref6,ref7}. Previous studies have also addressed regulatory direction and sign. For example, DGCGRN employs directed graph convolution to model regulatory direction \cite{ref8}, scSGL infers signed regulatory relationships from single-cell data \cite{ref9}, and DeepFGRN predicts directed activating and repressing relationships using bulk expression data \cite{ref10}.

Existing studies use gene expression, pseudotime, and regulatory network topology in different ways, but their effective integration for signed directed GRN inference remains an important question. Some graph-based approaches exploit recorded regulatory structure without explicitly retaining pseudotime ordering in gene representations, whereas pseudotime-based approaches use ordered expression changes to estimate regulatory relationships without necessarily distinguishing both regulatory direction and activating or repressing effects for ordered gene pairs. With incomplete reference GRNs, a key methodological challenge is to incorporate expression information from successive pseudotime stages into graph representations that preserve the direction and sign of recorded regulatory edges, thereby supporting predictions for candidate relationships not recorded in the reference network. In addition, restricting candidate regulatory sources to predefined transcription-factor lists can exclude other genes with potential regulatory roles within the selected gene set.

Incomplete reference GRNs may omit regulatory relationships originating from genes without previously annotated outgoing edges. Accordingly, we do not restrict candidate regulatory sources to predefined TF lists and formulate signed directed GRN inference as a three-class prediction task for ordered gene pairs in the selected gene set: No recorded edge, Activation, and Repression, with No recorded edge denoting missing reference annotation. The resulting predictions can also be compared across cellular trajectories to identify candidate regulatory gains, losses, and sign changes.

We developed PSD-GRN for signed directed GRN inference by integrating gene expression features from consecutive pseudotime stages with regulatory information from an incomplete prior GRN. At each stage, spectral convolution based on the signed directed magnetic Laplacian \cite{ref11} combines expression features with the direction and sign of recorded training edges encoded in a prior graph shared across pseudotime stages. Stage-residual connections integrate the resulting gene representations in pseudotime order for regulatory-state prediction of ordered gene pairs. All genes in the selected set can serve as candidate regulatory sources, without restriction to predefined transcription-factor lists. We evaluated PSD-GRN on directed edge-existence and three-class signed relationship prediction benchmarks and applied it to repair- and disease-associated epithelial trajectories in fibrotic lung to examine candidate regulatory changes.

\subsection{Main contributions}\label{main-contributions}

The main contributions of this study are summarized as follows:

\textbf{1.} We develop PSD-GRN, integrating pseudotime-stage expression features, stage-residual connections, and signed directed magnetic-Laplacian convolution to incorporate cellular-state changes and signed directed regulatory topology into graph representation learning.

\textbf{2.} PSD-GRN predicts regulatory states for ordered gene pairs without restricting candidate sources to predefined transcription factors, enabling the prioritization of relationships involving genes with no annotated outgoing edges in the reference GRN. Benchmark comparisons and ablation experiments demonstrate the predictive value of the proposed framework.

\textbf{3.} We applied PSD-GRN to compare candidate regulatory changes between repair- and disease-associated epithelial trajectories in fibrotic lung. The analysis identified candidate regulatory gains, losses, and sign changes, with a subset reproduced in an independent IPF cohort. Replication was significantly enriched among SPEN-originating relationships.

\section{Related Work}\label{related-work}

\subsection{Evolution of Topological Modeling in Single-Cell GRN Inference}\label{evolution-of-topological-modeling-in-single-cell-grn-inference}

Expression-based GRN inference methods estimate candidate regulatory relationships from gene expression profiles. GENIE3 uses tree-based regression to assess the contribution of candidate regulators to target-gene expression \cite{ref12}. CNNC represents the joint expression distribution of a gene pair as a two-dimensional histogram and learns interaction patterns using a convolutional neural network \cite{ref5}. Neither method explicitly incorporates reference GRN topology.

Graph-based methods exploit network topology to infer regulatory relationships. GRGNN examines local subgraphs surrounding candidate pairs \cite{ref13}, while GENELink aggregates gene-neighborhood information from an incomplete reference GRN \cite{ref6}. GNNLink combines expression features and network structure through an interaction graph encoder \cite{ref7}, and GCLink employs graph contrastive learning to improve inference when known regulatory edges are limited \cite{ref14}. AutoGERN learns adaptive edge-level representations during message passing \cite{ref15}, whereas GRACE incorporates causal mechanisms into graph-based regulatory inference \cite{ref16}.

Directed graph models account for the asymmetric roles of regulators and targets. DGCGRN uses first-order and separate second-order in- and out-neighborhood proximity matrices for directed graph convolution \cite{ref8}. DMDGRN uses a phase-based directed Laplacian to encode edge direction, graph augmentation to mitigate network sparsity, and multilayer directed convolution with residual connections to capture higher-order structure \cite{ref17}.

Regulatory sign distinguishes activating and repressing interactions. PoLoBag estimates positive and negative directed edge weights using polynomial Lasso ensemble regression \cite{ref18}. scSGL infers signed, undirected regulatory networks from differences in expression smoothness associated with positive and negative edges \cite{ref9}. DeepFGRN learns source- and target-specific neighborhood representations and combines them with bulk expression features to classify directed gene pairs as activation, repression, or no regulation \cite{ref10}.

Candidate-regulator selection also determines the scope of GRN inference. Methods such as GENELink \cite{ref6}, GCLink \cite{ref14}, and ATFGRN \cite{ref19} formulate regulatory prediction around candidate TF--target pairs, with candidate regulatory sources drawn from predefined TF sets. Such TF-focused prediction settings narrow the candidate space but leave genes outside these lists unevaluated as potential regulatory sources.

\subsection{Application of Pseudotime Information to Single-Cell GRN Inference}\label{application-of-pseudotime-information-to-single-cell-grn-inference}

Monocle reconstructs cellular trajectories from expression differences among sampled cells and assigns pseudotime to represent their relative ordering during development or differentiation \cite{ref20}. This inferred ordering provides a basis for examining gene-expression changes across cellular states without longitudinally tracking individual cells.

Pseudotime-based methods use ordered expression patterns to infer gene associations or candidate regulatory relationships. LEAP constructs gene co-expression networks using lagged correlations along pseudotime \cite{ref2}. SCODE models gene-expression dynamics using linear ordinary differential equations \cite{ref3}. SCRIBE applies restricted directed information to estimate directional dependence between the preceding expression state of a candidate regulator and the current state of a target gene \cite{ref21}. SINGE employs kernelized Granger causal regression to analyze irregularly sampled pseudotime expression profiles and combines results through ensembling \cite{ref4}.

Pseudotime provides an inferred ordering of cells rather than longitudinal observations of the same cells. SCRIBE reported poorer network inference from pseudotime than from true time-series data \cite{ref21}, while SINGE noted that insufficient or low-quality pseudotime information can impair network reconstruction \cite{ref4}. The BEELINE benchmark found that methods not requiring pseudotime generally performed better in its evaluated GRN inference settings \cite{ref22}. These findings indicate that the usefulness of pseudotime for GRN inference depends on trajectory quality and how the inferred ordering is incorporated into the model.

\section{Materials and Methods}\label{materials-and-methods}

\subsection{Problem Formulation}\label{problem-formulation}

Let \(X\in\mathbb{R}^{n\times m}\) denote the single-cell gene-expression matrix for \(n\) genes and \(m\) cells, with \(X_{ic}\) representing the expression of gene \(v_i\) in cell \(c\). The inferred pseudotime is \(\boldsymbol{\tau}=(\tau_1,\ldots,\tau_m)\). We construct a signed directed prior graph \(G=(V,E^+,E^-)\) from regulatory interactions recorded in a reference GRN, where \(V=\{v_1,\ldots,v_n\}\) is the selected gene set and \(E^+\) and \(E^-\) denote the recorded activating and repressive edge sets, respectively, with \(E^+\cap E^-=\varnothing\). The ordered pair \((v_i,v_j)\) denotes a candidate regulatory relationship from source gene \(v_i\) to target gene \(v_j\). Its signed directed adjacency matrix \(S\in\{-1,0,1\}^{n\times n}\) is defined as
\begin{equation}\label{eq:signed-adjacency}
S_{ij}=\begin{cases}
1, & (v_i,v_j)\in E^+,\\
-1, & (v_i,v_j)\in E^-,\\
0, & \text{otherwise}.
\end{cases}
\end{equation}

Given \(X\), \(\boldsymbol{\tau}\), and \(S\), PSD-GRN predicts a class-probability vector \(\mathbf p_{ij}\) for each ordered gene pair \((v_i,v_j)\) with \(i\ne j\). Directed edge-existence prediction distinguishes No recorded edge from Edge exists; signed directed prediction distinguishes No recorded edge, Activation, and Repression. No recorded edge denotes the absence of a corresponding relationship in the available regulatory annotations and does not establish biological nonregulation. Every gene in \(V\) may serve as a candidate regulatory source, without restriction to a predefined transcription-factor list.

For different cellular trajectories, the predicted states of the same ordered gene pairs can be compared to identify candidate changes in regulatory relationships. The cross-trajectory analysis is described in Section III-C.

\subsection{PSD-GRN Framework}\label{psd-grn-framework}

PSD-GRN comprises pseudotime stage-feature construction, signed directed magnetic-Laplacian convolution with stage-residual integration, and ordered gene-pair prediction (Fig. 1). These modules integrate gene expression information across pseudotime stages with the direction and sign of regulatory edges in the training prior graph.

\begin{figure*}[!t]
\centering
\includegraphics[width=\textwidth,height=5.40in,keepaspectratio]{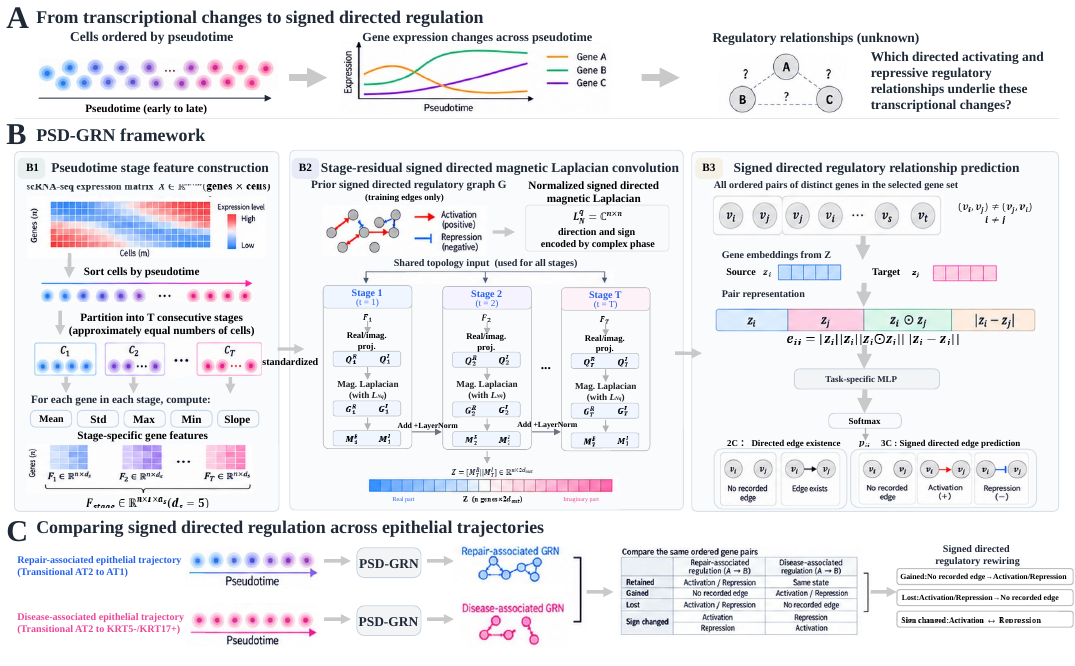}
\caption{Overview of PSD-GRN. (A) Task formulation: inferring directed activating and repressive regulatory relationships from pseudotime-ordered expression changes. (B) The PSD-GRN framework: B1, pseudotime stage feature construction. B2, stage-residual signed directed magnetic Laplacian convolution over a training prior graph shared across pseudotime stages. B3, classification of each ordered gene pair as No recorded edge, Activation, or Repression. (C) Cross-trajectory comparison design: PSD-GRN is trained separately on the repair-associated and disease-associated trajectories, and differences in the predicted state of the same ordered pair define gained, lost, and sign-changed relationships.}\label{fig:1}
\end{figure*}

\subsubsection{Pseudotime-Enhanced Node Feature Construction}\label{pseudotime-enhanced-node-feature-construction}

Gene expression is transformed by \(\log_{2}(x + 1)\), and pseudotime is min--max normalized to \(\lbrack 0,1\rbrack\) within each trajectory. Cells are sorted by pseudotime and partitioned into \(T\) consecutive stages \(C_{1},\ldots,C_{T}\) containing approximately equal numbers of cells. For gene \(v_{i}\) in stage \(C_{t}\), five expression statistics are extracted:

\begin{equation}\label{eq:stage-expression-features}
p_{it} = \left\lbrack \operatorname{mean}_{it},\operatorname{std}_{it},\max_{it},\min_{it},\operatorname{slope}_{it} \right\rbrack,
\end{equation}

\noindent where \(\operatorname{slope}_{it}\) is the coefficient of a univariate regression of expression on pseudotime within the stage. The first four statistics summarize local expression levels and variability, while the slope captures the local expression trend. Stacking gene features gives \(F_{t} \in \mathbb{R}^{n \times 5}\). Preserving stage order yields \(F_{stage} \in \mathbb{R}^{n \times T \times 5}\). Each stage-feature dimension is standardized across genes before convolution. Equal-sized stages provide comparable sample counts for estimating successive transcriptional states.

\subsubsection{Signed Directed Magnetic Laplacian}\label{signed-directed-magnetic-laplacian}

The training prior adjacency \(S\) is mapped to a Hermitian spectral operator through complex phases. Let \(A = |S|\), \(A_{s} = \left( A + A^{\mathsf{T}} \right)/2\), and \(D_{s} = \operatorname{diag}\left( A_{s}\mathbf{1} \right)\). With magnetic phase parameter \(q\), define

\begin{equation}\label{eq:phase-matrix}
B_{uv}^{q} = S_{uv}e^{iq} + S_{vu}e^{- iq},\quad\quad P_{uv}^{q} = \frac{B_{uv}^{q}}{\left| B_{uv}^{q} \right| + \varepsilon},
\end{equation}

\noindent where \(\varepsilon = 10^{- 5}\) ensures numerical stability. For a unidirectional regulatory interaction, reversing its direction conjugates the complex phase factor, while changing its sign multiplies that factor by \(-1\), equivalent to a phase shift of \(\pi\). The Hermitian adjacency and normalized signed directed magnetic Laplacian are then defined as

\begin{equation}\label{eq:normalized-magnetic-laplacian}
H^{q} = A_{s} \odot P^{q},\quad\quad L_{N}^{q} = I - D_{s}^{- 1/2}H^{q}D_{s}^{- 1/2},
\end{equation}

\noindent where \(\odot\) denotes element-wise multiplication. For zero-degree nodes, the diagonal factor in \(D_{s}^{- 1/2}\) is set to one. The conjugate symmetry \(P_{vu}^{q} = \overline{P_{uv}^{q}}\) and bound \(\left| P_{uv}^{q} \right| \leq 1\) give a Hermitian, positive-semidefinite \(L_{N}^{q}\) with real eigenvalues in \(\lbrack 0,2\rbrack\) \cite{ref11}.

\subsubsection{Spectral Convolution and Stage-Residual Integration}\label{spectral-convolution-and-stage-residual-integration}

Spectral filtering over \(L_{N}^{q}\) combines stage expression features with the training prior graph. Because \(L_{N}^{q}\) is Hermitian, it admits \(L_{N}^{q} = U\Lambda U^{\ast}\), and spectral convolution can be approximated without explicit eigendecomposition using Chebyshev polynomials \cite{ref23}:

\begin{equation}\label{eq:chebyshev-filter}
g_{\theta}*F \approx \sum_{k = 0}^{K}\theta_{k}T_{k}\left( {\widetilde{L}}_{N}^{q} \right)F,\quad\quad{\widetilde{L}}_{N}^{q} = \frac{2L_{N}^{q}}{\lambda_{\max}} - I,
\end{equation}

\noindent where \(\lambda_{\max} = 2\) and \(T_{k}\) denotes the \(k\)th Chebyshev polynomial.

At stage \(t\), separate learnable projections give \(Q_{t}^{R} = \operatorname{ReLU}\left( F_{t}W_{t}^{R} + b_{t}^{R} \right)\) and \(Q_{t}^{I} = \operatorname{ReLU}\left( F_{t}W_{t}^{I} + b_{t}^{I} \right)\). Writing \(Q_{t} = Q_{t}^{R} + iQ_{t}^{I}\), the Chebyshev responses are computed recursively as

\begin{equation}\label{eq:chebyshev-recursion}
\begin{matrix}
P_{t}^{(0)} & = Q_{t},\quad\quad P_{t}^{(1)} = {\widetilde{L}}_{N}^{q}Q_{t}, \\
P_{t}^{(k)} & = 2{\widetilde{L}}_{N}^{q}P_{t}^{(k - 1)} - P_{t}^{(k - 2)},\quad k \geq 2.
\end{matrix}
\end{equation}

The real and imaginary components of these Chebyshev responses are then transformed separately:

\begin{equation}\label{eq:complex-convolution}
\begin{matrix}
G_{t}^{R} & = \varphi\left( \sum_{k = 0}^{K}\operatorname{Re}\left( P_{t}^{(k)} \right)W_{t,k}^{R} + c_{t}^{R} \right), \\
G_{t}^{I} & = \varphi\left( \sum_{k = 0}^{K}\operatorname{Im}\left( P_{t}^{(k)} \right)W_{t,k}^{I} + c_{t}^{I} \right).
\end{matrix}
\end{equation}

\noindent where \(\varphi = \operatorname{Dropout} \circ \operatorname{ReLU} \circ \operatorname{LayerNorm}\), and \(c_{t}^{R},c_{t}^{I}\) are the real and imaginary convolution biases. The training prior graph and its spectral operator are shared across stages, while each stage applies one convolution block with stage-specific projection and convolution parameters.

To integrate successive stages, residual connections \cite{ref24} update the real and imaginary representations independently. For \(a \in \{ R,I\}\), \(M_{1}^{a} = G_{1}^{a}\) and

\begin{equation}\label{eq:stage-residual}
M_{t}^{a} = \operatorname{LayerNorm}\left( G_{t}^{a} + M_{t - 1}^{a} \right),\quad t = 2,\ldots,T.
\end{equation}

The final gene representation \(Z = \left\lbrack M_{T}^{R}\| M_{T}^{I} \right\rbrack\) combines expression features accumulated in pseudotime order with signed directed regulatory information encoded by the spectral operator.

\subsubsection{Ordered Gene-Pair Prediction}\label{ordered-gene-pair-prediction}

For an ordered pair \(\left( v_{i},v_{j} \right)\), the node embeddings \(z_{i}\) and \(z_{j}\) form

\begin{equation}\label{eq:pair-feature}
e_{ij} = \left\lbrack z_{i}\| z_{j}\|\left( z_{i} \odot z_{j} \right)\|\left| z_{i} - z_{j} \right| \right\rbrack,
\end{equation}

\noindent where \(\|\) is concatenation and \(\odot\) is the Hadamard product. The ordered concatenation of source and target embeddings allows the classifier to distinguish the two possible regulatory directions. The product and absolute difference describe feature interactions and contrasts. A task-specific multilayer perceptron produces class probabilities \(\mathbf{p}_{ij} = \operatorname{softmax}\left( \operatorname{MLP}\left( e_{ij} \right) \right)\).

\subsubsection{Training Objective}\label{training-objective}

Class-weighted Focal Loss \cite{ref25} is used for both prediction tasks, with class weights determined by the training-set label frequencies:

\begin{equation}\label{eq:focal-loss}
\mathcal{L} = - \frac{1}{\left| D_{train} \right|}\sum_{(i,j) \in D_{train}}^{}\alpha_{y_{ij}}\left( 1 - p_{ij,y_{ij}} \right)^{\gamma}\log p_{ij,y_{ij}},
\end{equation}

\noindent where \(p_{ij,y_{ij}}\) is the probability assigned to the labeled class \(y_{ij}\). The class weights are proportional to \(\sqrt{\left| D_{train} \right|/\left( Cn_{c} \right)}\), with \(C\) classes and \(n_{c}\) training instances in class \(c\), and are normalized to unit mean within each training split. Specific parameter settings and the benchmark protocol are reported in Section IV-B.

\subsection{Cross-Trajectory Regulatory Rewiring Analysis}\label{cross-trajectory-regulatory-rewiring-analysis}

To compare regulatory relationships across cell-state trajectories, PSD-GRN is fitted separately to each trajectory using a common gene set, the same reference regulatory network, and matched graph splits and initialization seeds. For each ordered gene pair, five independently trained models each predict one of three states: No recorded edge, Activation, or Repression. A trajectory-specific state is considered stable when at least four of the five models predict the same class. Only pairs with stable states in both trajectories are retained for cross-trajectory comparison. With changes defined from the repair-associated to the disease-associated trajectory, transitions from No recorded edge to Activation or Repression indicate candidate regulatory gains. The reverse transitions indicate losses. Activation--Repression switches indicate sign changes.

Candidate transitions are restricted to ordered pairs absent from the recorded reference network and ranked by a score combining transition-specific probability differences, model agreement, and confidence in the assigned states. Biological prioritization is then applied through source- and target-based eligibility criteria. These criteria are applied after prediction and do not restrict the ordered gene pairs evaluated by PSD-GRN. The selected candidate set is frozen before external evaluation. The score definition and selection rules are provided in the Supplementary Methods.

In an independent cohort, the corresponding trajectories are reconstructed and PSD-GRN is fitted anew, with frozen candidate pairs withheld from labeled supervision. For each trajectory, external regulatory states are assigned by selecting the class with the highest probability averaged over five independently trained models. Exact replication requires both trajectory-specific external states to match their corresponding discovery states for the same ordered pair. Permutation tests assess the overall replication count. Cohort-specific procedures, thresholds, and secondary analyses are described in Section V and the Supplementary Methods.

\section{Experiments}\label{experiments}

We evaluated PSD-GRN through three complementary questions: how accurately it predicts directed edge existence and signed regulatory states across single-cell datasets, how its key architectural components affect predictive performance, and how sensitive the results are to pseudotime stage resolution.

\subsection{Benchmark Datasets and Task Settings}\label{benchmark-datasets-and-task-settings}

To assess PSD-GRN across diverse cellular systems and gene-set sizes, we used seven single-cell RNA-seq datasets from the BEELINE benchmark \cite{ref22}: hESC, hHep, mDC, mESC, and three mouse hematopoietic stem cell datasets (mHSC-E, mHSC-GM, and mHSC-L). Pseudotime values were obtained from BEELINE, and gene-expression preprocessing followed GENELink \cite{ref6}. After low-expression filtering, up to 500 or 1,000 highly variable non-transcription-factor genes were combined with retained transcription factors to construct the TF+500 and TF+1000 gene sets. Transcription-factor membership was used only for gene-set construction. It was not provided as a model input or used to restrict predicted regulatory sources.

We constructed dataset-specific signed directed reference GRNs using regulatory interactions from TRRUST \cite{ref26} and DoRothEA \cite{ref27}. Activating and repressing interactions were labeled as positive and negative edges, respectively, with duplicate and conflicting records removed. Ordered gene pairs without a recorded interaction in the queried direction were eligible for sampling as \emph{No recorded edge}, indicating absent reference annotation rather than confirmed biological nonregulation. Training, validation, and test pairs were mutually disjoint within each split, and held-out regulatory edges were excluded from the training prior graph. Supplementary Table S1 summarizes the resulting networks. The mHSC-L dataset was excluded from signed prediction because its limited number of repressive edges prevented reliable stratified evaluation.

Two complementary tasks were used to evaluate signed directed GRN inference. For directed edge-existence prediction, recorded Activation and Repression edges were combined into a single regulatory-edge class and distinguished from sampled \emph{No recorded edge} pairs at a 1:1 ratio. This task assessed the identification of regulatory relationships in a specified direction. Signed directed relationship prediction further distinguished \emph{No recorded edge}, Activation, and Repression for each ordered gene pair, evaluating the model's ability to identify regulatory signs. For this three-class task, \emph{No recorded edge} pairs were sampled to match the combined number of recorded Activation and Repression edges within each split.

\subsection{Baselines and Implementation Details}\label{baselines-and-implementation-details}

For directed edge-existence prediction, we compared PSD-GRN with five baselines: GENELink, GNNLink, CNNC, AutoGERN, and GRACE. For signed directed relationship prediction, we evaluated DeepFGRN and GRACE adapted for three-class classification. PoLoBag estimates signed weights for directed regulatory edges. Its original evaluation emphasizes sign recovery among recorded edges, and applying it to the present three-class task requires an additional rule for determining edge existence. scSGL learns an undirected signed network and does not distinguish opposite regulatory directions. We therefore used DeepFGRN, whose prediction objective is closer to the present task, and a three-class adaptation of GRACE as the graph-based baseline. All methods were evaluated using identical benchmark gene sets, training, validation, and test pairs, including the same sampled negative pairs, and task-specific evaluation metrics. These comparisons assessed both the identification of directed regulatory relationships and the discrimination of their activating and repressing signs.

For each dataset and task, five data splits were generated with an approximate 80\%/10\%/10\% training/validation/test ratio, retaining connectivity-preserving edges in the training graph. Model parameters were selected through preliminary iterative trials. PSD-GRN used five pseudotime stages, a hidden dimension of 64, Chebyshev order \(K = 1\), magnetic phase \(q = 0.1\pi\), dropout of 0.5, and Focal Loss with \(\gamma = 0.75\). Training used Adam \cite{ref28} (initial learning rate \(10^{- 3}\), weight decay 0.05) with cosine annealing for up to 1,000 epochs. Validation was performed every 25 epochs, with early stopping after 20 consecutive evaluations without improvement. Model selection was based on validation loss for edge-existence prediction and validation Macro-F1 for signed prediction. AUROC, AUPRC, Macro-F1, and Micro-F1 were reported for both tasks, with multiclass MCC additionally reported for signed prediction. Results were averaged over five splits.

\begin{figure*}[!t]
\centering
\includegraphics[width=\textwidth,height=6.72in,keepaspectratio]{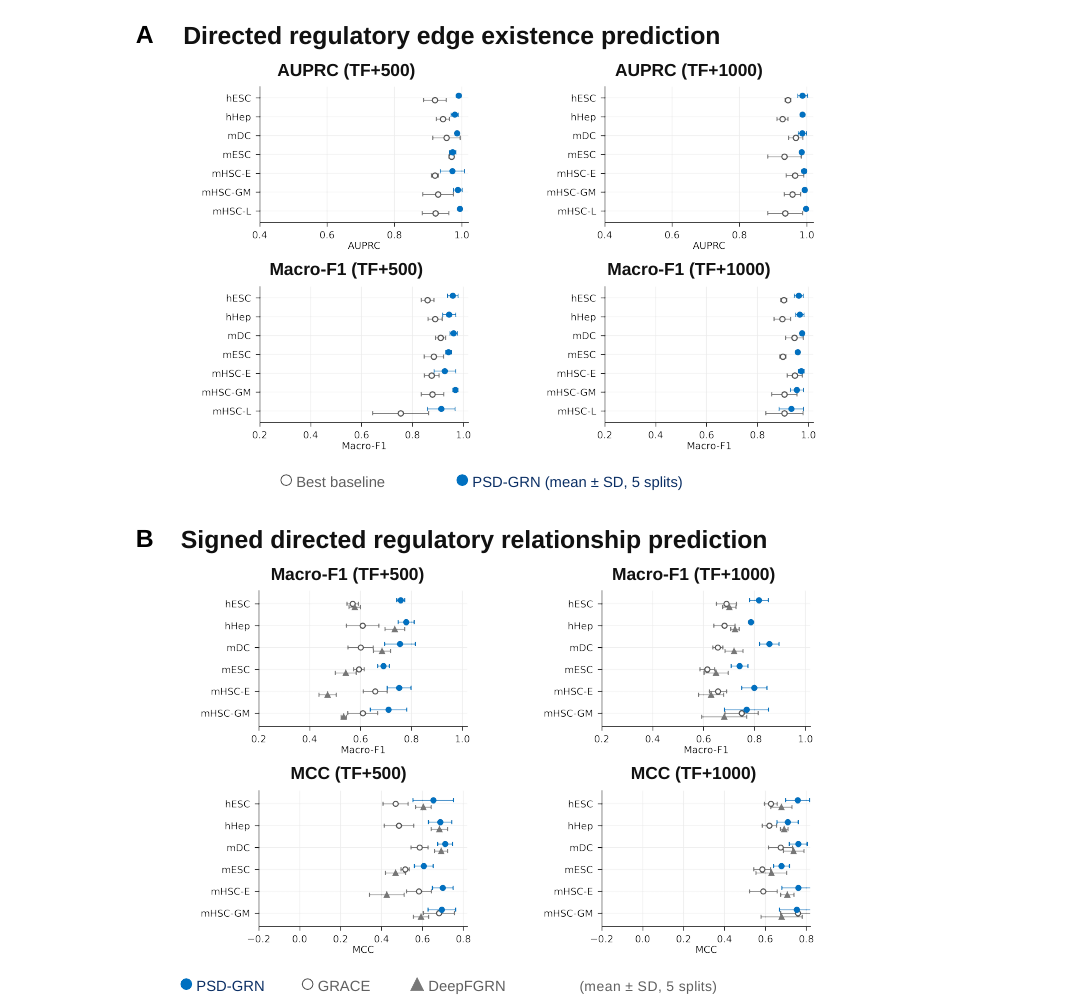}
\caption{Benchmark performance. (A) Directed edge-existence prediction: per-dataset AUPRC and Macro-F1 of PSD-GRN (filled circles) and the best-performing baseline for each dataset and metric under each gene-set setting (open circles). (B) Signed directed relationship prediction: per-dataset Macro-F1 and MCC of PSD-GRN, GRACE, and DeepFGRN. Values are means over five splits. Error bars denote standard deviations.}\label{fig:2}
\end{figure*}

\subsection{Performance Comparison}\label{performance-comparison}

For directed edge-existence prediction, PSD-GRN achieved the highest AUROC, AUPRC, Macro-F1, and Micro-F1 across all seven datasets under both TF+500 and TF+1000 settings (Fig. 2A, Table I, and Supplementary Table S2). Under TF+500, the cross-dataset mean AUROC and AUPRC reached 98.40\% and 98.33\%, compared with 92.83\% and 91.79\% for GRACE, the strongest baseline in the cross-dataset averages (Table I). Under TF+1000, the corresponding values increased to 99.05\% and 98.99\%, with the advantage maintained across all seven datasets. These results demonstrate consistent discrimination between reference-annotated directed regulatory edges and sampled unrecorded gene pairs across the evaluated cellular systems and gene-set sizes.

\begin{table*}[!t]
\centering
\footnotesize
\caption{Cross-dataset mean \ensuremath{\pm} SD (\%) of directed edge-existence prediction performance on seven datasets under TF+500 and TF+1000. Within each dataset, performance was first averaged over five data splits. SD describes variation across the seven dataset-level means. Best means for each gene-set setting and metric are shown in bold. Detailed per-dataset mean \ensuremath{\pm} SD over splits is reported in Supplementary Table S2.}\label{tab:1}
\setlength{\tabcolsep}{4.3pt}
\begin{tabular}{@{}llcccc@{}}
\toprule
\textbf{Gene set} & \textbf{Model} & \textbf{AUROC} & \textbf{AUPRC} & \textbf{Macro-F1} & \textbf{Micro-F1} \\ \midrule
TF+500 & \textbf{PSD-GRN} & \textbf{98.40} \(\pm\) \textbf{0.73} & \textbf{98.33} \(\pm\) \textbf{0.82} & \textbf{94.47} \(\pm\) \textbf{1.84} & \textbf{94.50} \(\pm\) \textbf{1.80}  \\
TF+500 & GENELink & 85.42 \(\pm\) 5.52 & 86.87 \(\pm\) 3.96 & 82.90 \(\pm\) 8.51 & 83.67 \(\pm\) 6.96  \\
TF+500 & GRACE & 92.83 \(\pm\) 0.98 & 91.79 \(\pm\) 1.10 & 85.54 \(\pm\) 6.80 & 85.87 \(\pm\) 6.06  \\
TF+500 & CNNC & 81.59 \(\pm\) 4.33 & 82.19 \(\pm\) 4.66 & 72.90 \(\pm\) 3.60 & 73.17 \(\pm\) 3.60  \\
TF+500 & AutoGERN & 88.04 \(\pm\) 11.57 & 88.40 \(\pm\) 10.04 & 75.51 \(\pm\) 13.00 & 77.46 \(\pm\) 10.85  \\
TF+500 & GNNLink & 75.39 \(\pm\) 14.46 & 77.42 \(\pm\) 11.12 & 68.32 \(\pm\) 13.39 & 69.44 \(\pm\) 12.53  \\
TF+1000 & \textbf{PSD-GRN} & \textbf{99.05} \(\pm\) \textbf{0.38} & \textbf{98.99} \(\pm\) \textbf{0.42} & \textbf{96.02} \(\pm\) \textbf{1.31} & \textbf{96.03} \(\pm\) \textbf{1.29}  \\
TF+1000 & GENELink & 88.07 \(\pm\) 6.00 & 89.90 \(\pm\) 3.50 & 84.52 \(\pm\) 10.57 & 85.28 \(\pm\) 9.22  \\
TF+1000 & GRACE & 95.11 \(\pm\) 1.62 & 94.45 \(\pm\) 1.88 & 91.42 \(\pm\) 1.87 & 91.45 \(\pm\) 1.87  \\
TF+1000 & CNNC & 83.83 \(\pm\) 3.86 & 83.92 \(\pm\) 4.13 & 76.73 \(\pm\) 3.76 & 76.88 \(\pm\) 3.81  \\
TF+1000 & AutoGERN & 89.81 \(\pm\) 6.48 & 89.50 \(\pm\) 5.77 & 74.77 \(\pm\) 10.10 & 76.63 \(\pm\) 8.43  \\
TF+1000 & GNNLink & 76.16 \(\pm\) 16.88 & 78.28 \(\pm\) 12.01 & 70.86 \(\pm\) 10.09 & 72.24 \(\pm\) 8.81  \\
\bottomrule
\end{tabular}
\end{table*}

Signed directed relationship prediction requires distinguishing \emph{No recorded edge}, Activation, and Repression for each ordered gene pair. We therefore emphasized Macro-F1 and multiclass MCC: Macro-F1 evaluates the three regulatory states with equal class weight, while MCC measures overall classification agreement under class imbalance. PSD-GRN achieved the highest cross-dataset mean performance on all five metrics under both gene-set settings (Fig. 2B, Table II, and Supplementary Table S3). Its mean Macro-F1 increased from 74.04\% under TF+500 to 79.51\% under TF+1000, while MCC increased from 67.52\% to 73.64\%. Under TF+1000, DeepFGRN achieved 68.41\% Macro-F1 and 68.72\% MCC (Table II). PSD-GRN also attained the highest Macro-F1 on all six datasets under both settings, demonstrating consistent advantages in classifying direction-specific regulatory states.

\begin{table*}[!t]
\centering
\footnotesize
\caption{Cross-dataset mean \ensuremath{\pm} SD (\%) of three-class signed directed relationship prediction on six datasets under TF+500 and TF+1000. Within each dataset, performance was first averaged over five data splits. SD describes variation across the six dataset-level means. Best means for each gene-set setting and metric are shown in bold. Detailed per-dataset mean \ensuremath{\pm} SD over splits is reported in Supplementary Table S3.}\label{tab:2}
\setlength{\tabcolsep}{4.3pt}
\begin{tabular}{@{}llccccc@{}}
\toprule
\textbf{Gene set} & \textbf{Model} & \textbf{AUROC} & \textbf{AUPRC} & \textbf{Macro-F1} & \textbf{Micro-F1} & \textbf{MCC} \\ \midrule
TF+500 & \textbf{PSD-GRN} & \textbf{90.75} \(\pm\) \textbf{3.66} & \textbf{76.25} \(\pm\) \textbf{3.23} & \textbf{74.04} \(\pm\) \textbf{3.06} & \textbf{80.08} \(\pm\) \textbf{2.32} & \textbf{67.52} \(\pm\) \textbf{3.55}  \\
TF+500 & GRACE & 81.73 \(\pm\) 1.75 & 65.39 \(\pm\) 2.82 & 60.61 \(\pm\) 2.65 & 73.57 \(\pm\) 4.85 & 55.32 \(\pm\) 7.24  \\
TF+500 & DeepFGRN & 85.07 \(\pm\) 4.75 & 68.41 \(\pm\) 6.46 & 59.04 \(\pm\) 9.11 & 75.62 \(\pm\) 5.78 & 57.77 \(\pm\) 10.01  \\
TF+1000 & \textbf{PSD-GRN} & \textbf{94.84} \(\pm\) \textbf{0.93} & \textbf{81.31} \(\pm\) \textbf{2.31} & \textbf{79.51} \(\pm\) \textbf{3.70} & \textbf{84.04} \(\pm\) \textbf{2.17} & \textbf{73.64} \(\pm\) \textbf{3.17}  \\
TF+1000 & GRACE & 86.14 \(\pm\) 2.20 & 69.75 \(\pm\) 2.69 & 67.47 \(\pm\) 4.11 & 78.92 \(\pm\) 3.92 & 64.20 \(\pm\) 5.99  \\
TF+1000 & DeepFGRN & 90.60 \(\pm\) 2.32 & 76.47 \(\pm\) 4.30 & 68.41 \(\pm\) 3.49 & 81.77 \(\pm\) 1.96 & 68.72 \(\pm\) 3.30  \\
\bottomrule
\end{tabular}
\end{table*}

PSD-GRN allows all genes in the selected gene set to serve as potential regulatory sources. We therefore examined TCF7L1 in the hHep TF+1000 dataset as a representative case. Although TCF7L1 was included in the gene set, it was absent from the predefined TF list and had no recorded outgoing regulatory relationships in the reference GRN. PSD-GRN identified 45 high-confidence activating and 17 repressing candidate relationships originating from TCF7L1. Table S13 presents 16 of these high-confidence candidate relationships alongside their external evidence assessments (Supplementary Methods 1.8).

\subsection{Ablation Study}\label{ablation-study}

We evaluated four ablation variants to examine the contributions of pseudotime representation and signed directed graph learning to three-class regulatory prediction. Experiments were conducted on six datasets under the TF+1000 setting (Fig. 3 and Supplementary Table S4). Three variants examined the pseudotime-related components: w/o SR replaced inter-stage residual connections with concatenation. The w/o PT variant randomly reassigned cell-expression profiles to fixed pseudotime positions using a shared permutation. The w/o PEF variant replaced stage-feature construction with a dimension-matched MLP. The w/o MLC variant replaced magnetic Laplacian convolution with a standard GCN to evaluate the contribution of the joint signed-directed topology encoding. All variants used the same data splits and sampling rules as the full model.

All four ablations reduced predictive performance (Fig. 3B,C and Supplementary Table S4). Mean Macro-F1 decreased from 79.51\% for the full model to 75.9\% without stage residual connections, 74.3\% with disrupted pseudotime ordering, and 70.7\% when the stage-feature extractor was replaced by an MLP. These results support the contributions of cellular-state ordering, stage-specific feature construction, and residual integration. Replacing magnetic Laplacian convolution with a standard GCN produced the largest performance decline, with mean AUROC decreasing from 94.84\% to 82.4\% and Macro-F1 from 79.51\% to 55.7\%. This substantial decline highlights the contribution of magnetic Laplacian-based convolution to learning signed directed regulatory relationships.

\begin{figure*}[!t]
\centering
\includegraphics[width=\textwidth,height=4.75in,keepaspectratio]{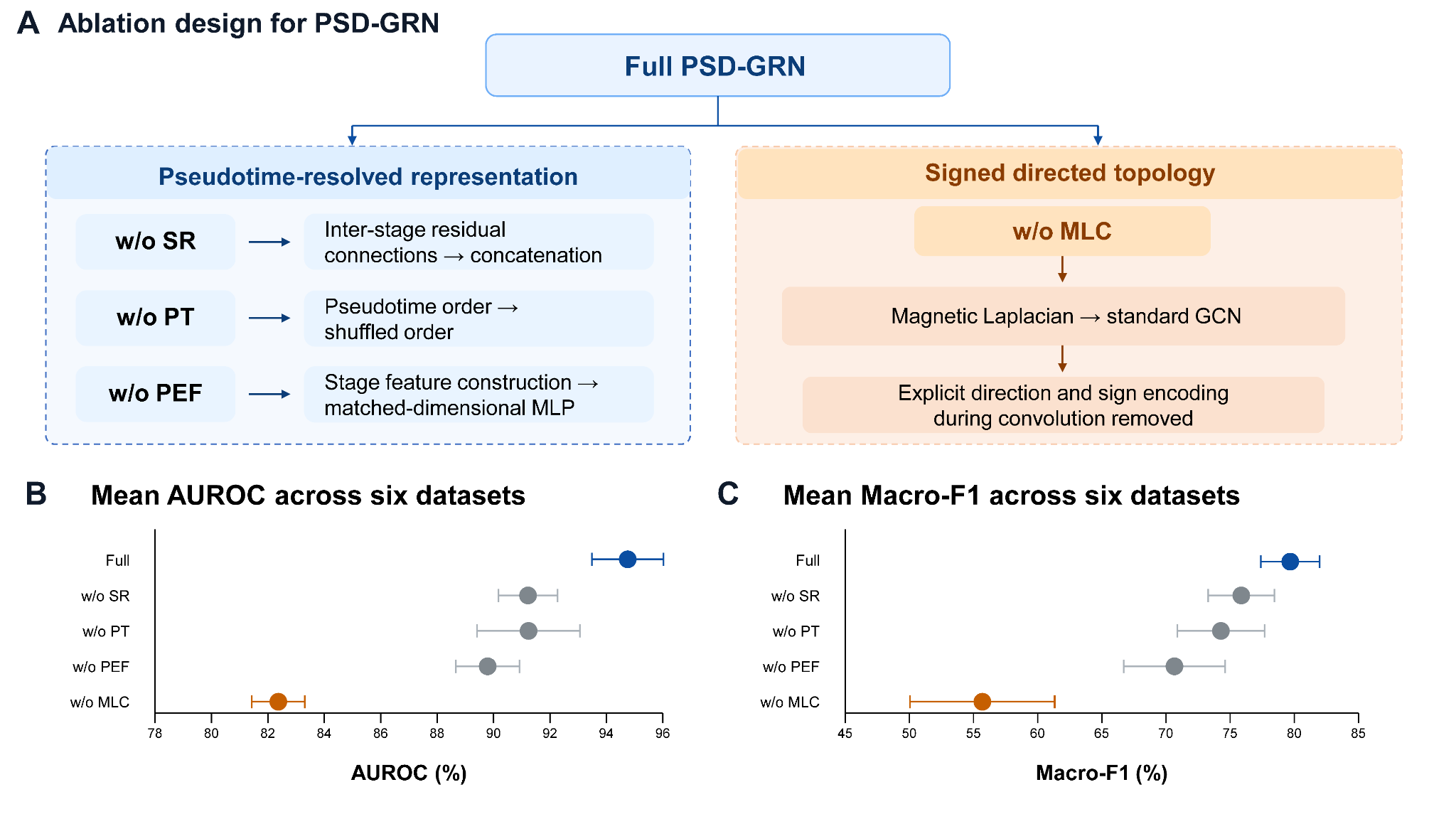}
\caption{Ablation of PSD-GRN on signed relationship prediction under TF+1000. (A) Ablation design: four variants target the pseudotime-resolved representation (w/o SR, w/o PT, w/o PEF) or the signed directed topology encoding (w/o MLC). (B, C) Mean AUROC (B) and Macro-F1 (C) across the six signed-task datasets. Error bars denote standard deviations across datasets. w/o SR: inter-stage residual connections replaced by concatenation. w/o PT: pseudotime ordering replaced by a shared shuffled cell order. w/o PEF: stage feature construction replaced by a matched-dimensional MLP. w/o MLC: magnetic Laplacian convolution replaced by a standard GCN.}\label{fig:3}
\end{figure*}

\subsection{Sensitivity to Pseudotime Stage Resolution}\label{sensitivity-to-pseudotime-stage-resolution}

The number of pseudotime stages determines the resolution of stage-wise expression features, potentially affecting the balance between capturing cellular-state changes and retaining sufficient cells per stage. We therefore evaluated \(T = 3\) to \(7\) on the six signed prediction datasets under TF+1000, keeping all other settings fixed. Across these resolutions, mean AUROC ranged from 94.28\% to 94.84\%, and Macro-F1 from 78.05\% to 79.51\%, indicating modest performance variation (Supplementary Fig. S1). Both metrics reached their highest values at \(T = 5\), supporting its use in the main experiments.

\section{Application to Pulmonary Fibrosis}\label{application-to-pulmonary-fibrosis}

\subsection{Study Design and Epithelial Trajectories}\label{study-design-and-epithelial-trajectories}

Building on previously characterized epithelial populations in pulmonary fibrosis \cite{ref29,ref30}, we reconstructed two trajectories from Transitional AT2 cells toward AT1 cells (repair-associated) and KRT5\ensuremath{-}/KRT17+ cells (disease-associated). The discovery cohort GSE135893 included IPF and other fibrotic/interstitial lung diseases, while IPF samples from the independent GSE136831 dataset were used for replication. Following Seurat/SCTransform preprocessing \cite{ref31,ref32}, the trajectories were reconstructed separately using Slingshot \cite{ref33}, with Transitional AT2 as the starting state. In GSE135893, the repair-associated and disease-associated trajectories comprised 922 and 1,095 cells, respectively, including 623 shared Transitional AT2 cells. This shared starting population provided a common cellular reference for comparing the two inferred epithelial trajectories.

To make cross-trajectory comparisons less sensitive to differences in model configuration, PSD-GRN was trained separately on each trajectory using the same gene set and reference regulatory network, identical training, validation, and test splits of candidate gene pairs, and matched initialization seeds. Five models were trained independently for each trajectory. To assess reproducibility in an independent cohort, the discovery candidates were frozen before evaluation in GSE136831, where the corresponding trajectories were reconstructed and PSD-GRN was retrained with the frozen pairs excluded from labeled supervision. Detailed procedures are provided in the Supplementary Methods.

\subsection{Distinct Regulatory Rewiring between Divergent Epithelial Trajectories}\label{distinct-regulatory-rewiring-between-divergent-epithelial-trajectories}

To identify regulatory-state changes associated with divergent epithelial trajectories, we compared stable predictions for the same ordered gene pairs between the repair- and disease-associated trajectories in GSE135893 (Fig. 4). We focused on state transitions involving pairs absent from the reference GRN and ranked them using transition-specific probability differences, model agreement, and prediction confidence. Two complementary post-prediction criteria prioritized relationships involving eligible regulatory sources or established epithelial/fibrosis-associated marker targets. Together, these criteria yielded 464 unique candidates for subsequent evaluation (Supplementary Methods and Table S5).

Of the 464 candidates, 286 (61.6\%) represented predicted disease-associated regulatory gains, 155 (33.4\%) represented losses, and 23 (5.0\%) represented Activation--Repression switches (Fig. 4B,C). To examine whether the inferred regulatory rewiring primarily reflected gene expression differences between the two trajectories, we assessed the association between rewiring scores and stage-wise expression divergence across the 464 candidates. No significant correlation was observed (Spearman \ensuremath{\rho} = 0.011, P = 0.818). To assess sensitivity to donor composition, we performed five donor-balanced paired resamplings, in which 49 candidates (10.6\%) reproduced their complete predicted state transitions in at least four runs (Supplementary Table S6).

\begin{figure*}[!t]
\centering
\includegraphics[width=\textwidth,height=5.15in,keepaspectratio]{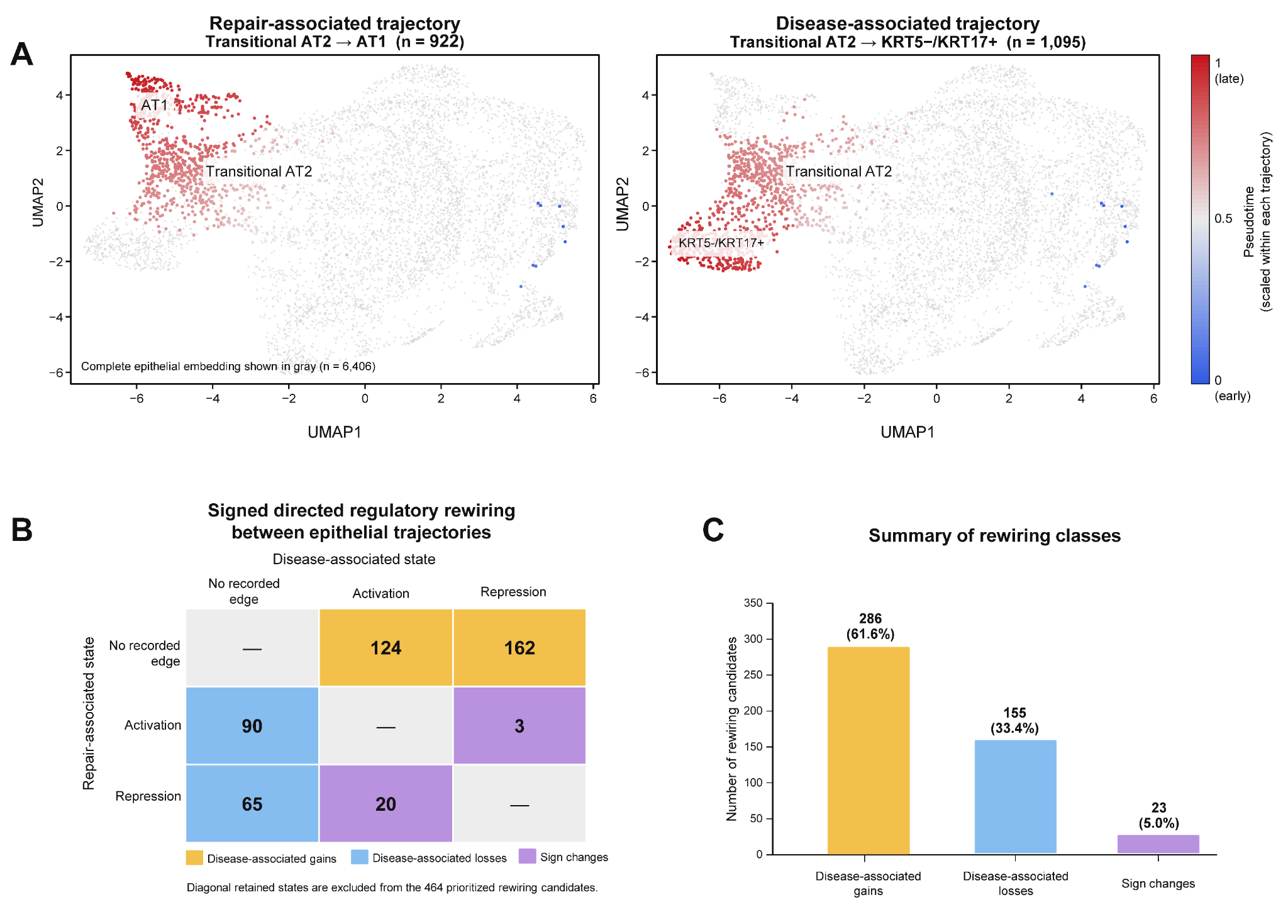}
\caption{Trajectory-specific signed directed regulatory rewiring in fibrotic lung. (A) UMAP visualization of the repair-associated Transitional AT2\(\rightarrow\)AT1 trajectory (922 cells) and the disease-associated Transitional AT2\(\rightarrow\)KRT5\(-\)/KRT17+ trajectory (1,095 cells) in GSE135893. The complete epithelial embedding is shown in gray, and pseudotime is scaled within each trajectory. (B) Cross-trajectory regulatory-state transitions among the 464 frozen discovery candidates. (C) Composition of the candidate set: 286 disease-associated gains, 155 losses, and 23 sign changes.}\label{fig:4}
\end{figure*}

\subsection{Independent Replication Identifies Reproducible SPEN-Centered Rewiring}\label{independent-replication-identifies-reproducible-spen-centered-rewiring}

The 464 discovery candidates were evaluated in the independent IPF cohort GSE136831, comprising 209 cells from 28 donors in the repair-associated trajectory and 485 cells from 30 donors in the disease-associated trajectory. After matching the gene space, 463 candidates remained evaluable. Based on class probabilities averaged across five independently trained models per trajectory, 58 candidates (12.5\%) exactly reproduced both trajectory-specific discovery states (Fig. 5B). A permutation test preserving the frequencies of the six discovery transition types yielded an average of 27.5 exact matches (95\% empirical interval, 19--36). None of 100,000 permutations reached the observed count (empirical P = \(1.0 \times 10^{-5}\)). The result remained significant under a source-preserving permutation (empirical P = 0.0028). Among the 58 ensemble-exact candidates, 25 matched both discovery states in at least three of the five paired external model runs (Supplementary Table S7).

To determine whether the replicated regulatory rewiring was concentrated around particular regulatory sources, we performed an exploratory source-level enrichment analysis across 128 candidate regulators. SPEN was the only source significantly enriched for exact replication after multiple-testing correction, with 8 of 10 SPEN-originating relationships reproduced compared with 50 of 453 relationships from other sources (odds ratio = 32.24, one-sided Fisher's exact P = \(1.42 \times 10^{-6}\), BH q = \(1.82 \times 10^{-4}\), Fig. 5C). Because regulatory gains showed higher overall replication rates and all ten SPEN candidates were disease-associated gains, we repeated the enrichment analysis within gain candidates. SPEN remained significant (8/10 versus 39/275, odds ratio = 24.21, BH q = \(7.75 \times 10^{-4}\)). The eight replicated relationships comprised three Activation gains and five Repression gains (Fig. 5D and Supplementary Table S9).

\begin{figure*}[!t]
\centering
\includegraphics[width=\textwidth,height=5.10in,keepaspectratio]{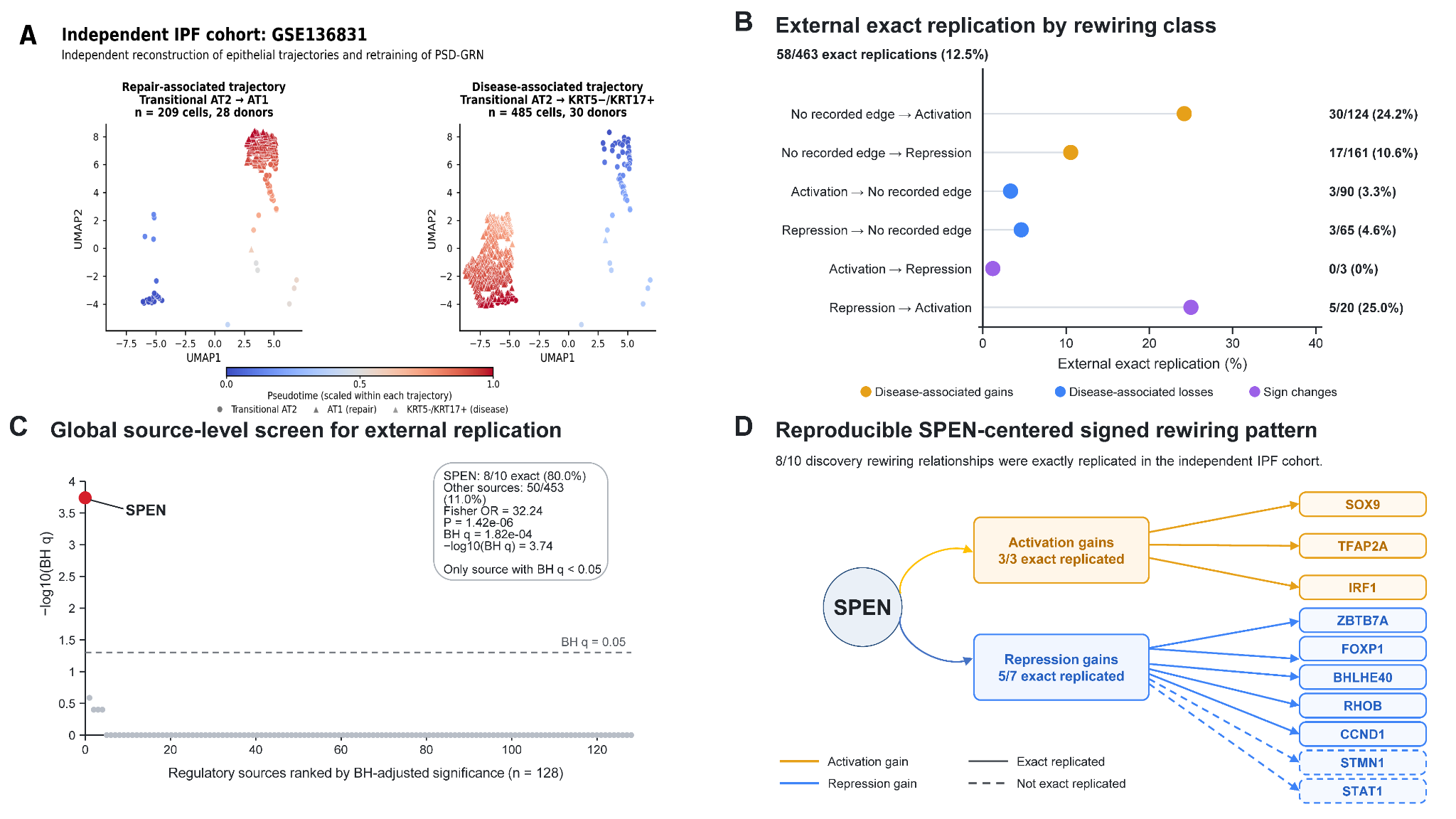}
\caption{Independent IPF replication identifies reproducible SPEN-centered regulatory rewiring. (A) Independently reconstructed trajectories in GSE136831 (209 cells from 28 donors for repair and 485 cells from 30 donors for disease). (B) Exact external replication of the frozen candidates by rewiring class. (C) Source-level enrichment across 128 regulatory sources. SPEN was the only source with BH \(q < 0.05\) (8/10 exact replications versus 50/453 for other sources, odds ratio \(= 32.24\)). (D) SPEN-centered discovery rewiring relationships. Solid and dashed arrows denote relationships exactly replicated and not exactly replicated in the independent cohort, respectively.}\label{fig:5}
\end{figure*}

\section{Discussion}\label{discussion}

Across benchmark datasets, PSD-GRN improved directed edge-existence prediction and three-class signed regulatory relationship prediction, with particularly clear gains in Macro-F1 and MCC for the signed task. The ablation results help explain these improvements. Disrupting pseudotime ordering reduced performance even when the stage-wise feature computation was retained, indicating that cellular-state ordering contributes predictive information. Replacing magnetic Laplacian convolution with a standard GCN produced the largest performance decline, highlighting the contribution of magnetic spectral convolution to signed directed regulatory prediction. Moreover, performance varied modestly across three to seven pseudotime stages, suggesting that the results were not critically dependent on a single stage resolution.

Beyond benchmark prediction, the fibrotic lung application illustrates the potential of PSD-GRN for comparing signed directed regulatory states across divergent cellular trajectories. The two trajectories shared a Transitional AT2 starting population and were analyzed using matched gene pairs and model configurations, providing a consistent basis for identifying trajectory-associated differences in predicted regulatory states. Importantly, discovery candidates were fixed before external evaluation, and the models were independently retrained with these pairs excluded from labeled supervision. Using the exact-replication criterion requiring agreement in both trajectory-specific regulatory states, 58 of 463 candidates were reproduced in the independent IPF cohort, significantly exceeding expectations under both permutation controls. These reproducible predictions provide a focused set of candidate regulatory changes for further investigation into the regulatory mechanisms underlying divergent epithelial trajectories.

SPEN emerged as a focal regulatory source because its candidate gains showed enriched exact replication, including in the gain-only analysis. This concentration suggests a reproducible SPEN-associated component of predicted regulatory differences between epithelial trajectories. These findings prioritize SPEN-associated candidate relationships for future investigation in relevant lung epithelial models.

Several limitations warrant consideration. Pseudotime inferred from cross-sectional single-cell profiles depends on trajectory reconstruction and cell-state sampling. Reference GRN incompleteness and cohort differences may influence candidate selection and cross-dataset replication. Four candidates met both the donor-balanced discovery robustness and external exact-replication criteria (Supplementary Tables S6--S7). Exploratory RBPJ/SHARP chromatin \cite{ref34} and SOX9-knockdown analyses are reported in Supplementary Methods 1.6--1.7 and Tables S10–S11 but do not directly validate the predicted SPEN regulatory relationships. Future studies could assess additional cohorts and trajectory methods and directly test prioritized candidates in relevant lung epithelial models.

\section{Conclusion}\label{conclusion}

PSD-GRN integrates pseudotime-resolved expression features with signed directed spectral graph learning to predict the regulatory states of ordered gene pairs, including candidates missing from reference networks. The method achieved higher average performance than the strongest baselines in both directed edge-existence and signed relationship prediction, supporting the value of combining cellular-state ordering with regulatory topology. In fibrotic lung, comparing predictions for the same gene pairs across repair- and disease-associated epithelial trajectories revealed candidate gains, losses, and sign changes. A subset reproduced in an independent IPF cohort under a frozen-candidate design, with significant replication enrichment among SPEN-originating predictions. These findings offer a path from cellular trajectories to reproducible, testable hypotheses about disease-associated regulatory rewiring.

\textbf{Supplementary Material.} Supplementary Material S1 (PDF) contains the Supplementary Methods, Supplementary Figure S1, and legends for Supplementary Tables S1--S13. Supplementary Material S2 (XLSX) contains Supplementary Tables S1--S13.

\section*{Statements and Declarations}
Competing interests: The authors declare no competing interests.

Ethics approval: Not applicable. This study used previously published, publicly available, de-identified transcriptomic datasets and involved no recruitment of participants or collection of new human samples.

Consent to participate: Not applicable to this secondary analysis of publicly available, de-identified datasets. No participants were recruited or contacted in the present study.

Consent for publication: Not applicable. No identifiable participant information or images are presented in this manuscript.

Data availability: All datasets analyzed in this study are publicly available. Benchmark scRNA-seq datasets are available through GEO (GSE75748, GSE81252, GSE48968, GSE98664, and GSE81682). The benchmark preprocessing used the GENELink resources and pseudotime information from BEELINE. The fibrotic-lung discovery cohort is GSE135893, the independent IPF cohort is GSE136831, the RBPJ/SHARP chromatin dataset is GSE249973, and the SOX9 perturbation dataset is GSE284531. Recorded signed directed regulatory relationships were obtained from TRRUST and DoRothEA.

Code availability: The source code of PSD-GRN is publicly available at https://github.com/zbscott/psd-grn. Additional information required to reanalyze the data reported in this paper is available from the lead contact upon request.

Authors' contributions: Qi Chen: Methodology, Software, Formal analysis, Validation, Visualization, Investigation, Writing -- original draft, Writing -- review and editing. Kai Wang: Software, Data curation, Investigation, Validation, Writing -- review and editing. Hongliang Zhang: Methodology, Formal analysis, Validation, Writing -- review and editing. Wei Xiong: Project administration, Funding acquisition, Writing -- review and editing. Yuan-nong Ye: Funding acquisition, Resources, Supervision, Writing -- review and editing. Bin Zhao: Conceptualization, Methodology, Supervision, Project administration, Funding acquisition, Writing -- review and editing.




\begin{IEEEbiography}[{\includegraphics[width=1in,height=1.25in,keepaspectratio]{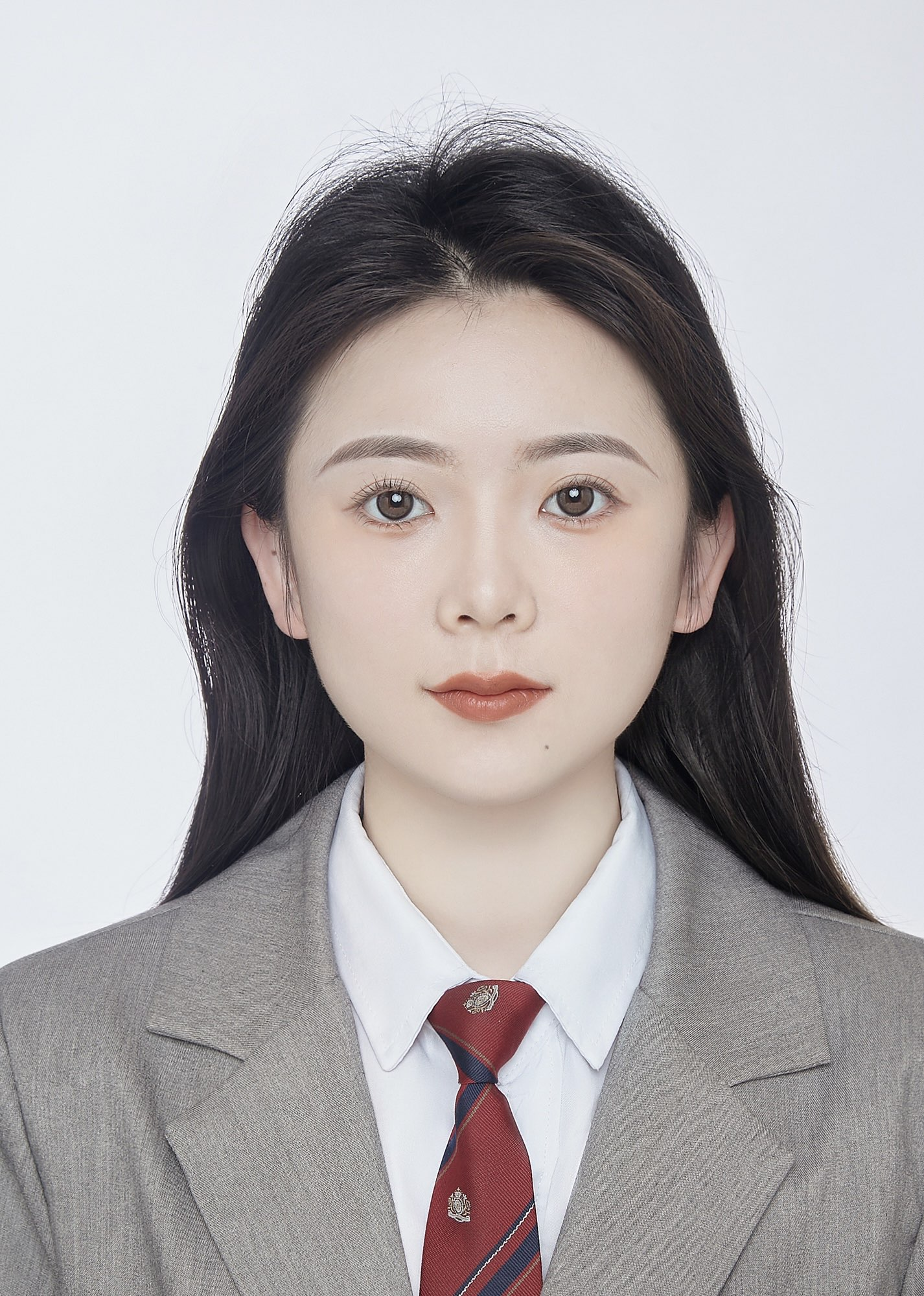}}]{Qi Chen}
is currently pursuing a master's degree in applied statistics at the School of Mathematics and System Science, Xinjiang University, China. Research interests include graph deep learning.
\end{IEEEbiography}

\begin{IEEEbiography}[{\includegraphics[width=1in,height=1.25in,keepaspectratio]{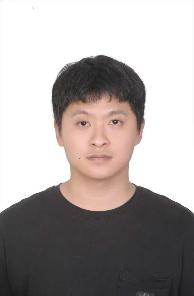}}]{Kai Wang}
is currently pursuing the master's degree in biomedical engineering at Guizhou Medical University, China. His research interests include bioinformatics.
\end{IEEEbiography}

\begin{IEEEbiography}[{\includegraphics[width=1in,height=1.25in,keepaspectratio]{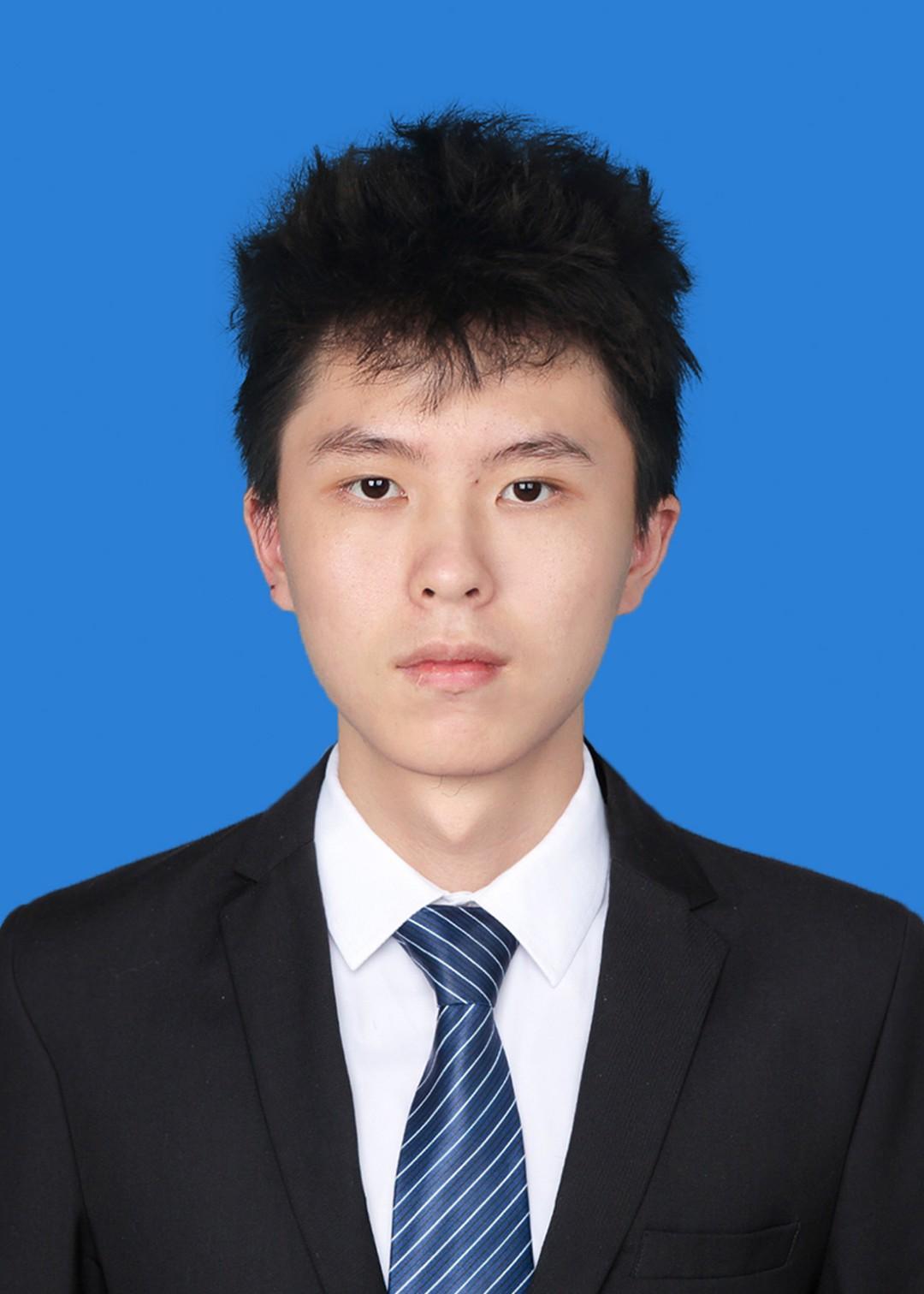}}]{Hongliang Zhang}
is currently pursuing a master's degree in applied statistics at the School of Mathematics and System Science, Xinjiang University, China. Research interests include graph deep learning.
\end{IEEEbiography}

\begin{IEEEbiography}[{\includegraphics[width=1in,height=1.25in,keepaspectratio]{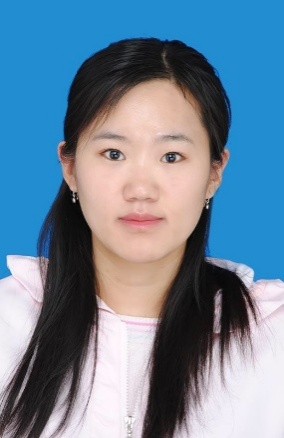}}]{Wei Xiong}
received the PhD degree from Xinjiang University, in 2014. She is currently an associate professor with the School of Mathematics and System Science, Xinjiang University. Her major research interests include graph theory and application.
\end{IEEEbiography}

\begin{IEEEbiography}[{\includegraphics[width=1in,height=1.25in,keepaspectratio]{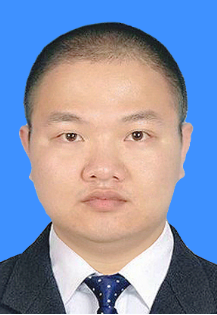}}]{Yuan-nong Ye}
received the PhD degree from University of Electronic Science and Technology of China in 2015. He is currently an associate professor with the School of Biology and Engineering, Guizhou Medical University. His major research interests include bioinformatics.
\end{IEEEbiography}

\begin{IEEEbiography}[{\includegraphics[width=1in,height=1.25in,keepaspectratio]{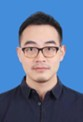}}]{Bin Zhao}
received the PhD degree from Xinjiang University, in 2016. He is currently an associate professor with the School of Biology and Engineering, Guizhou Medical University. His major research interests include deep learning.
\end{IEEEbiography}
\end{document}